\documentclass[12pt,preprint]{aastex}

\usepackage{epsfig} 

\newcommand{\beq}{\begin{equation}}
\newcommand{\eeq}{\end{equation}}
\newcommand{\beqa}{\begin{eqnarray}}
\newcommand{\eeqa}{\end{eqnarray}}
\newcommand{\om}{\Omega_m}

\newcommand{\ls}{\mathrel{\raise0.27ex\hbox{$<$}\kern-0.70em \lower0.71ex\hbox{{$\scriptstyle \sim$}}}}

\begin{document} 

\title{Snapping Supernovae at \lowercase{$z>1.7$}} 
\author{Greg Aldering, Alex G.\ Kim, Marek Kowalski, Eric V. Linder, 
Saul Perlmutter} 
\affil{Lawrence Berkeley National Laboratory, Berkeley, CA 94720} 

\begin{abstract}
We examine the utility of very high redshift Type Ia supernovae 
for cosmology and systematic uncertainty 
control.  Next generation space surveys such as the 
Supernova/Acceleration Probe (SNAP) will obtain thousands of supernovae 
at $z>1.7$, beyond the design redshift for which the supernovae will be 
exquisitely characterized.  We find that any $z\gtrsim2$ standard 
candles' use for cosmological parameter 
estimation is quite modest and subject to pitfalls; we examine 
gravitational lensing, redshift calibration, and contamination effects 
in some detail.  The very high redshift supernovae --- 
both thermonuclear and core collapse --- will provide 
copious interesting information on star formation, environment, and evolution.
However, the new observational systematics that must be faced, as well as the limited
expansion of SN-parameter space afforded, does not point to high value
for $1.7<z<3$ SNe~Ia in controlling evolutionary systematics relative to what SNAP can
already achieve at $z<1.7$.
Synergy with observations from JWST and 
thirty meter class telescopes afford rich opportunities for advances 
throughout astrophysics. 
\end{abstract} 



\section{Introduction \label{sec:intro}}

Supernovae play important roles in astrophysics and cosmology, from 
stellar nucleosynthesis and subsequent star formation, to injection 
of substantial energy as galaxies form and evolve, 
to production of neutron stars, black holes, and 
gamma ray bursts, neutrinos, and subsequent gravitational waves, to 
use as standardized measures of the expansion history of the universe 
and probes of dark energy.  

Their visible luminosities are so great they can be seen to high
redshifts, and certain classes (in particular Type~Ia) can be 
calibrated more accurately than
any other astronomical object to provide robust distance measures.
For these reasons, supernovae observations are pushing to higher and
higher redshifts, looking back over the majority of the history of the
universe.  Since they become difficult to observe from the ground as their
flux redshifts out of the optical window and the terrestrial sky
background grows very bright, space based measurements are necessary.
Wide field instruments are needed to achieve sufficient numbers
to study, and stringent, systematics controlled experiments to derive
accurate results.  For example, the Supernova/Acceleration Probe (SNAP:
\citet{aldering04}) is carefully designed specifically to achieve
high quality (in both statistics and systematics), well characterized, 
photometric and spectroscopic observations of thousands of supernovae.

Space extends the reach of a non-cryogenic telescope to 1.7$\mu$m, beyond
which thermal noise swamps the astronomical signal.  To this wavelength
limit the SiII 6300\AA\ spectral feature used for classification of
Type Ia supernovae (SNe~Ia) can be observed to redshift $z=1.7$.  This
redshift limit matches extremely well the optimum depth for dark energy
investigations, beyond which the sensitivity to the dark energy characteristics
fades away \citep{linhut03}.  However, a 2-m class space telescope such as
SNAP will observe large numbers of supernovae, both SNe~Ia and other types,
at redshifts $z>1.7$.  For example, the visible flux could be followed
out to $z\approx4$ and the near UV down to 2500\AA\ 
(SNe~Ia have almost no emission $< 2500$\AA) out to $z\approx6$.

In this article we investigate the usefulness of space based, wide
field observations of SNe beyond $z=1.7$ for cosmology and astrophysics.
Section \ref{sec:cos} addresses the cosmological impact of extending
precision distance measurements to $z>1.7$, including gravitational
lensing effects; this will be generally applicable to any standardized
candle, not just supernovae.  The rates and yield of supernovae of all 
types are
discussed in \S\ref{sec:measure}, together with measurement issues such
as light curve fitting, redshift determination, Malmquist bias, and
supernova typing.  Section~\ref{sec:sys} investigates what we can learn
about progenitor age, metallicity, and dust properties, and how this 
impacts treatment
of systematic uncertainties of the $z<1.7$ sample.  We summarize in
\S\ref{sec:concl} the prospects for using very high redshift SNe (obtained
for ``free'', and in conjunction with JWST or a TMT) to advance a variety
of astrophysical fields.

\section{Cosmology beyond \lowercase{$z=1.7$}} \label{sec:cos}

The discovery of the recent acceleration of the cosmic expansion is 
a breakthrough in the quest to understand the universe.  To reveal 
the nature of the dark energy responsible for the acceleration requires 
accurate measurements throughout the accelerating epoch and back 
into the time of deceleration.  However, indefinite extension through 
the matter dominated, deceleration epoch is of limited use (see, e.g., 
\citet{linhut03}), with diminished science returns for $z\gtrsim1.7$. 
This lack of leverage, together with increased uncertainty -- and 
biasing -- from photometric, redshift, 
and environment measurement errors and gravitational lensing (de)amplification 
of the source flux, is one of the flaws in seeking to push 
putative standardizable candles such as gamma ray bursts or gravitational 
waves to higher redshifts. 

\subsection{Dark energy} \label{sec:dark}

We begin by addressing this question in more detail: could there be 
some dark energy scenarios where percent level measurements of 
$z\gtrsim1.7$ distances are crucial in uncovering the nature of dark energy? 
On a coarse level the answer is immediately no; we know that structure 
formed in the universe, in a manner not too different from a universe 
dominated by matter at $z>2$, so dark energy cannot be dynamically 
important at high redshift (see \citet{early} for details about 
growth of structure constraints).  However it is worthwhile to address this 
question quantitatively, from the distance perspective, to show that even 
percent level distance measurements are insufficient. 

\begin{figure}[!ht]
\begin{center}
\psfig{file=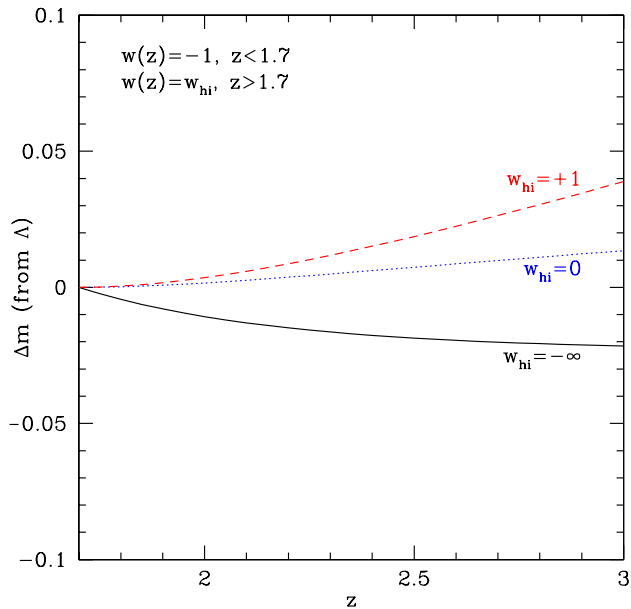, width=3.4in} 
\caption{Magnitude deviations above $z=1.7$ from a fiducial model 
remain small even for drastic deviations from the fiducial equation of 
state.  Here, at $z<1.7$ the model is assumed to be the 
fiducial cosmological constant, then jumps to an equation of state 
$w_{\rm hi}$ at higher redshifts. 
} 
\label{fig:dmlam}
\end{center} 
\end{figure}

Let us consider a model where future measurements of the distances 
out to $z=1.7$ have determined that in that range the dark energy acts 
like a cosmological constant.  We then ask what measurements are required 
at $z>1.7$ to see even drastically different behaviors at higher 
redshift.  In particular, suppose the dark energy equation of state 
$w(z)$ suddenly plunges at $z=1.7$ to $w_{\rm hi}=-\infty$, rises to 
act like matter, 
with $w_{\rm hi}=0$, or overshoots to the upper physical limit of 
$w_{\rm hi}=+1$.  Figure~\ref{fig:dmlam} shows the results in terms 
of the magnitude difference (0.0217 mag equals 1\% distance precision) 
from the model where the dark energy continues to behave as measured 
at $z<1.7$, i.e.\ a cosmological constant.  

\begin{figure}[!ht]
\begin{center}
\psfig{file=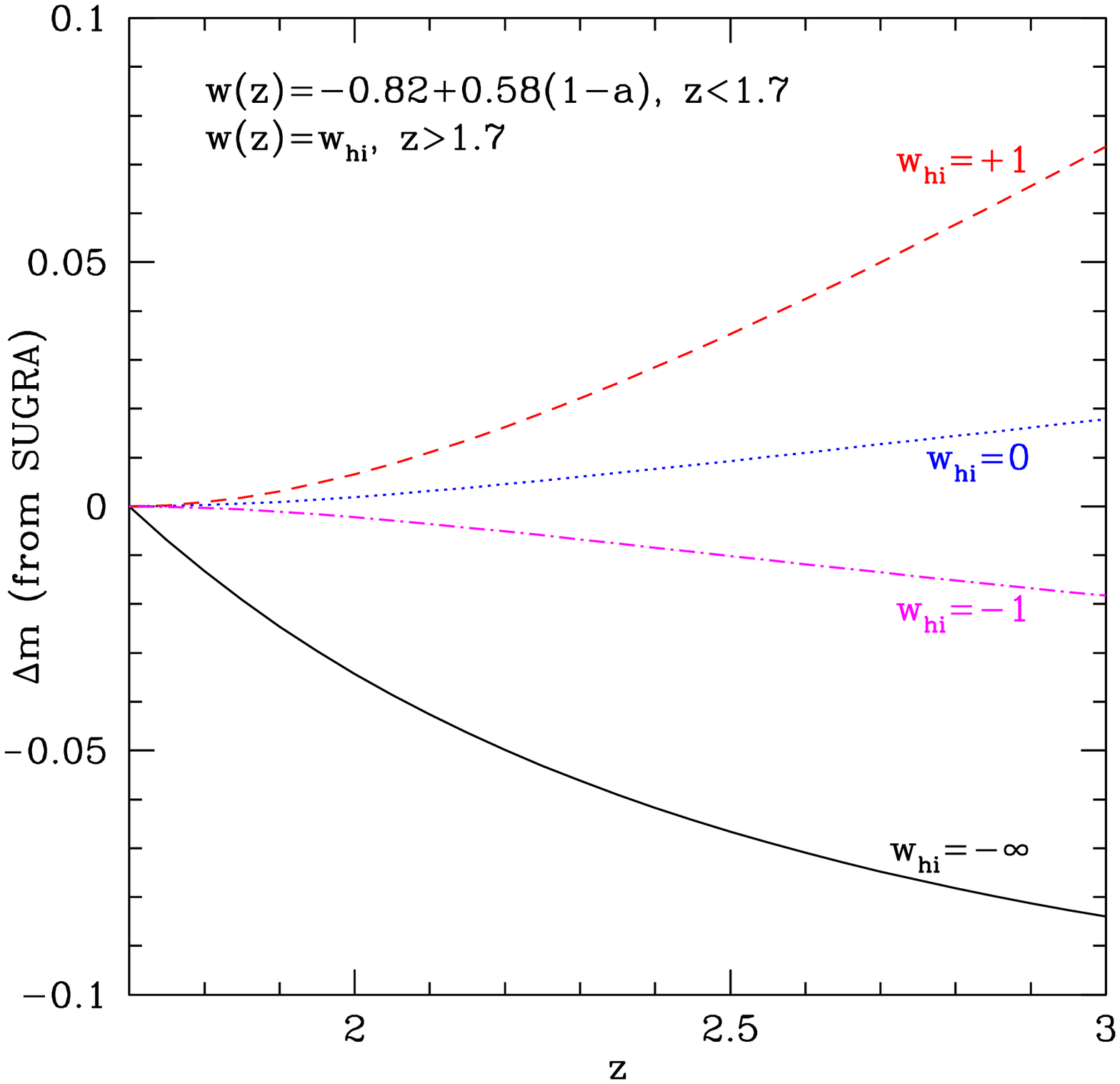, width=3.4in}
\caption{As Fig.~\ref{fig:dmlam}, but with a fiducial model of SUGRA. 
This has a less negative equation of state than the cosmological 
constant (near present observational bounds), giving greater persistence 
of dark energy density to higher redshifts.  Magnitude variations due to 
$z>1.7$ behavior still remain small, especially in the physically 
reasonable range $-1\le w_{\rm hi}\le0$. 
}
\label{fig:dmsug}
\end{center}
\end{figure}

Even if we allow the dark energy to suddenly have $w\ll-1$ or behave 
like matter, such extreme behavior could not be observed with 1\% 
distance measurements even if extended beyond $z=3$.  The extraordinary 
upper bound behavior of $w=+1$, which would upset structure formation, 
has less than a 2\% effect on distances to $z=3$.  Any reasonable dark 
energy evolution appearing at $z>1.7$ could not be realistically probed 
by extending accurate measurements beyond the fiducial $z=1.7$.  This 
conclusion holds as well for models with dynamics at lower redshifts, 
giving more persistent dark energy density at high redshifts. 
The SUGRA model at the upper limit of current observations, joined with 
extreme high redshift behavior of the equation of state lowering 
to the cosmological constant value or rising to the matter value, also 
does not show greater than 1\% distance deviations if mapped out to $z=3$ 
(see Fig.~\ref{fig:dmsug}). 

This demonstrates that even near perfect standard candles in a perfect 
universe do not motivate precision distance observations for $z>1.7$.  
As we will show, 1\% distance accuracy will be enormously difficult to 
approach at these high redshifts, due in large part to systematics.  
Distance markers with greater intrinsic dispersion than SNe~Ia, such 
as gamma ray bursts (which have 1000 times the intrinsic dispersion 
of SNe~Ia \citep{friedman05}), are even less useful.  Moreover, both the 
cosmology and observations have uncertainties and biasing mechanisms. 
We address some of these in the following sections, beginning with 
gravitational lensing due to the universe being imperfectly homogeneous. 

\subsection{Gravitational lensing} \label{sec:glens}

As light propagates over longer distances, the source has increased
probability of amplification or deamplification due to gravitational
scattering (lensing) off large scale structure.  Averaged over many
sources at a given redshift, the mean flux is unchanged but there is
increased dispersion of the source luminosity function.  Because the
amplification probability distribution is non-Gaussian (skewed toward 
deamplification), the received flux distribution is also asymmetric and 
incomplete 
sampling due to a finite number of sources will impose a bias.  This
presents a danger in drawing cosmology conclusions from relatively small
numbers of high redshift sources, hence affecting gamma ray bursts and
gravitational wave sources much more than supernovae.  We consider both
the dispersion and bias effects.

\citet{holzlin} give a rigorous approach to including lensing in 
cosmological distance measurements.  (Note that treating lensing through 
the matter power spectrum instead of the Monte Carlo method of 
\citet{holzlin} is only a rough approximation and does not incorporate 
non-Gaussianity or bias.)  Their prescription is that for more than 
$\sim10$ sources per $0.1z$ interval, the statistics approach Gaussian, 
the bias is much less than a statistical standard deviation, and the 
extra dispersion can be treated statistically by adding in quadrature 
to each supernovae an uncertainty 
\beq 
\sigma_{\rm lens}=0.1z/(1+0.07z). \label{eq:siglens} 
\eeq 
This approximation extends the \citet{holzlin} result of 
$\sigma_{\rm lens}=0.093z$ for dispersion in magnitudes to redshifts 
$z>2$ (also see \citet{premadimartel}).  

For the fiducial case of a SNAP-like SN survey, we expect 
more than a thousand SNe~Ia in the range $z=1.7-3$ (see \S\ref{sec:rates}) 
so treating lensing as an added dispersion is an excellent approximation, 
in contrast to cases where there might be only tens of sources in this 
range.  We include lensing effects in this manner for the remainder of 
this article.  Note that out to $z=1.7$, lensing has a small effect for a 
SNAP-like survey, with less than 5\% degradation of cosmological 
parameter estimation, for SN combined with the distance to the CMB last 
scattering surface. 
In \S\ref{sec:lenstype} we consider the tails of the lensing amplification 
distribution where 
lensing can cause appreciable (de)magnification, leading to possible 
confusion between SN types if categorized solely by luminosity.  In the 
next section we examine cosmological bias, due to lensing in part but 
mostly from misestimation of redshift. 

\subsection{Redshift errors and bias} \label{sec:zerr} 

The two key quantities entering the distance probe are the flux (as 
just discussed) and the redshift. An important question is how much 
the redshift uncertainty degrades the Hubble diagram and cosmological 
parameter estimation. 
The great majority of the thousands of SNe discovered at $z>1.7$, 
especially by a survey dedicated to the follow up of $z\le1.7$ events, 
will lack spectroscopic redshifts.  We examine the possibility of 
photometric redshifts in \S\ref{sec:photoz}; here we investigate the effects 
in terms of a generic uncertainty $\delta z$. 

Propagating redshift uncertainties through even a Fisher matrix approach 
is nontrivial due to correlations and dependencies.  A correct technique 
would be to include the full covariance matrix ($N_{\rm source}\times 
N_{\rm source}$) of magnitude errors induced by redshift uncertainties.  
However, the standard practice in the literature (e.g.\ \citet{hutkim}) 
is to consider only uncorrelated, Gaussian, redshift bin centroid errors.  
While \citet{hutkim} implement this in terms of new fit parameters, to 
be marginalized over, we adopt the approach of treating redshift 
uncertainties as an additional source of dispersion, i.e.\ 
approximating the full covariance matrix by diagonal entries. In addition, 
though, we also consider redshift errors as a systematic effect and, 
as a function of cosmological parameter bias tolerated, bound the 
allowed correlated error.

In the statistical approach, the uncertainty $(\partial m/\partial z)
\,\delta z$ is added in quadrature with the other magnitude errors.
The typical amplitude of the distance modulus contribution to the 
source magnitude uncertainty is 1.7-0.8 over the
range $z=1.7-3$.  This is not the full story, however, as the source 
magnitude must be corrected to the calibrated peak magnitude.  The redshift
uncertainty also propagates into this through,
e.g., the light curve width or stretch correction\footnote{While redshift
errors could also propagate into extinction and K-corrections, we do not
consider those here due to the following.  First, redshift misestimation
would be absorbed into a simultaneous fit of extinction parameters 
\citep{kimmiq}.  Second, K-corrections have a periodic structure in the
magnitude-redshift relation that does not mimic the effects of dark
energy \citep{linosc} and so does not degrade cosmological parameter
estimation given a long redshift baseline and a well-designed filter 
set.}, which contributes to $m$ a term $\alpha(s-1)$ \citep{p99}.
Since the stretch is defined as the observed width $s_0$ corrected for time
dilation, $s=s_0/(1+z)$, then a redshift error gives rise to a magnitude
error of $dm=\alpha ds=-\alpha s\,dz/(1+z)$.  We adopt the typical SN~Ia
values of $\alpha=1.5$ and $s=1$.

\begin{figure}[!ht]
\begin{center}
\psfig{file=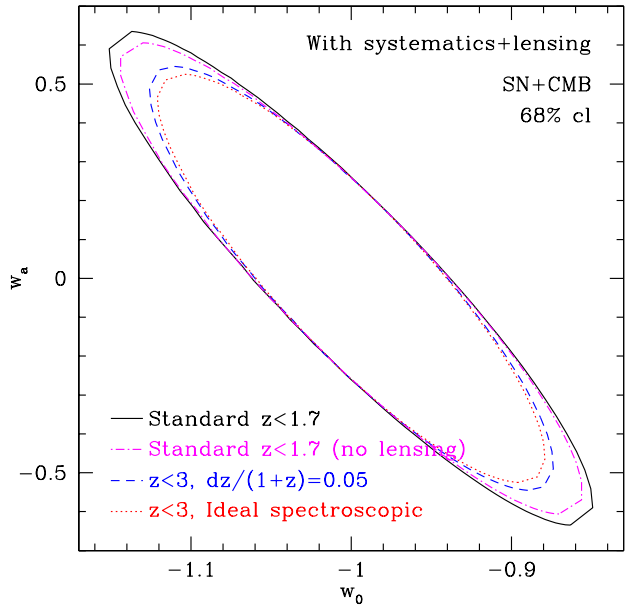, width=3.4in}
\caption{The standard, tightly systematics controlled, $z=1.7$ depth 
supernova survey is seen to be close to optimal for determination of 
the dark energy equation of state parameters.  Lensing adds less than 
5\% degradation.  Extending the depth to $z=3$, under the ideal case 
of full spectroscopy and tight systematics control, improves dark 
energy parameter estimation by less than 20\%. 
}
\label{fig:lenszerr}
\end{center}
\end{figure}

\begin{figure}[!ht]
\begin{center}
\psfig{file=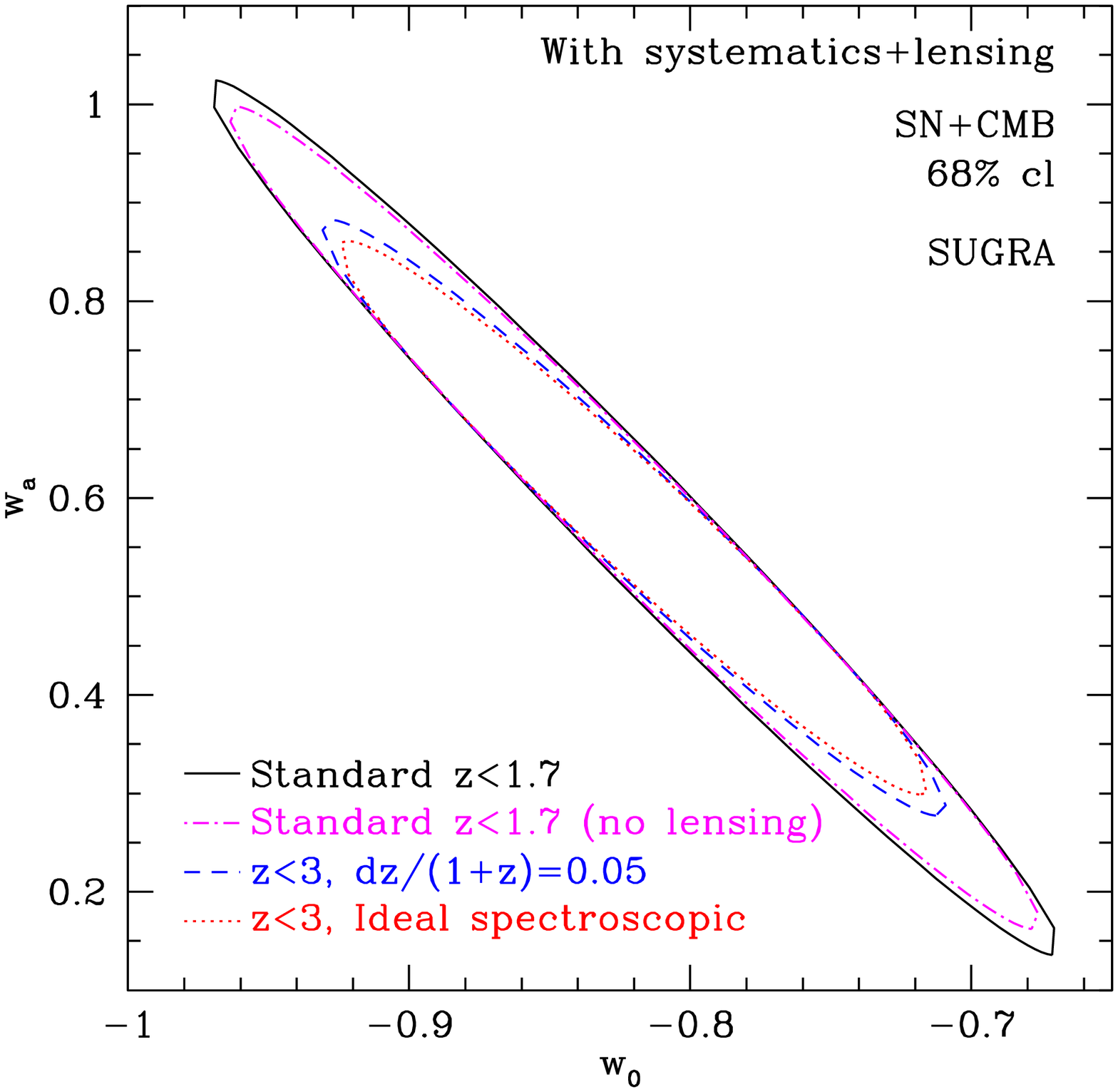, width=3.4in}
\caption{As Fig.~\ref{fig:lenszerr}, for SUGRA dark energy.  Due to 
the model's greater persistence of dark energy to higher redshift, 
the improvement of the ideal $z=3$ depth survey relative to the 
standard $z=1.7$ depth survey can be $\sim35\%$.  However, such 
persistence may be ruled out by structure growth constraints, e.g.\ 
SUGRA deviates by 40\% in the matter power spectrum amplitude 
relative to the cosmological constant case. 
}
\label{fig:lenszerrsug}
\end{center}
\end{figure}

Figure \ref{fig:lenszerr} shows the influence of the photometric redshift
statistical uncertainties (and gravitational lensing amplification)
on the dark energy equation of state (EOS) determination (as does
Fig.~\ref{fig:lenszerrsug} for the SUGRA case), assuming a SNAP-like
survey extended with 1000 SNe distributed uniformly in $z=1.7-3$.
The present value of the EOS is $w_0$ and its time variation is measured
by $w_a=-dw/da|_{z=1}$.  First, we see that as mentioned the effects
of lensing are small for the canonical SNAP-like SN survey to $z=1.7$.
Note that even if the space observations were able to continue detailed,
spectroscopic characterization of thousands of SNe~Ia out to $z=3$ 
(note that diffraction limited spectroscopy time increases as $(1+z)^6$), 
with the same systematic error floor of $dm_{\rm sys}=0.02(1+z)/2.7$, the 
improvement in cosmological parameter estimation is modest.

Next we consider no spectroscopy for SNe~Ia in the range $z=1.7-3$, yet
somehow keeping the same systematic floor, however adding new systematics
from photometric redshift errors.  These take the form of random Gaussian
uncertainties $\sigma_{\ln(1+z)}$ on the individual source redshifts.
Even an uncertainty of 0.05, readily achieved today for $z\lesssim1$ 
\citep{ilbert}, brings the cosmology estimation contour most of
the way to the ideal case of full spectroscopy.  However, the caveat of
retaining all other systematics at the same level, without spectroscopic
characterization of the SN, is a major obstacle.  Even if this were
possible, the improvement on extending the distance measurement 
photometrically to $z=3$ relative to the standard $z=1.7$ survey is only 
15\% (less than 20\% with full spectroscopy).  This confirms
that the $z=1.7$ depth is well chosen.

A more challenging problem is constraining a systematic component of 
the photometric redshift errors of the $z>1.7$ SNe, rather than their random 
scatter.  A systematic offset 
will lead to a bias in cosmological parameter extraction, of 
more concern than an increased dispersion.  Here the redshift error 
is treated as a coherent systematic shift $\Delta z$ of the derived 
photometric redshift relative to the true redshift.  The resulting 
bias on a cosmological parameter $p_i$ can be obtained through the 
standard Fisher bias 
\beq
\Delta p_i=F^{-1}{}_{ij}\int_0^{z_{\rm max}} dz\,N(z)\frac{\partial 
m}{\partial p_j}\Delta m(z)/\sigma_m^2(z), 
\eeq 
where $\Delta m=(dm/dz)\,\Delta z$ is the magnitude offset (including 
stretch) caused by 
the redshift systematic $\Delta z$, $\sigma_m$ is the overall 
magnitude dispersion, $N(z)$ is the SN redshift distribution, and 
$F_{ij}$ is the Fisher matrix. 

\begin{figure}[!ht]
\begin{center}
\psfig{file=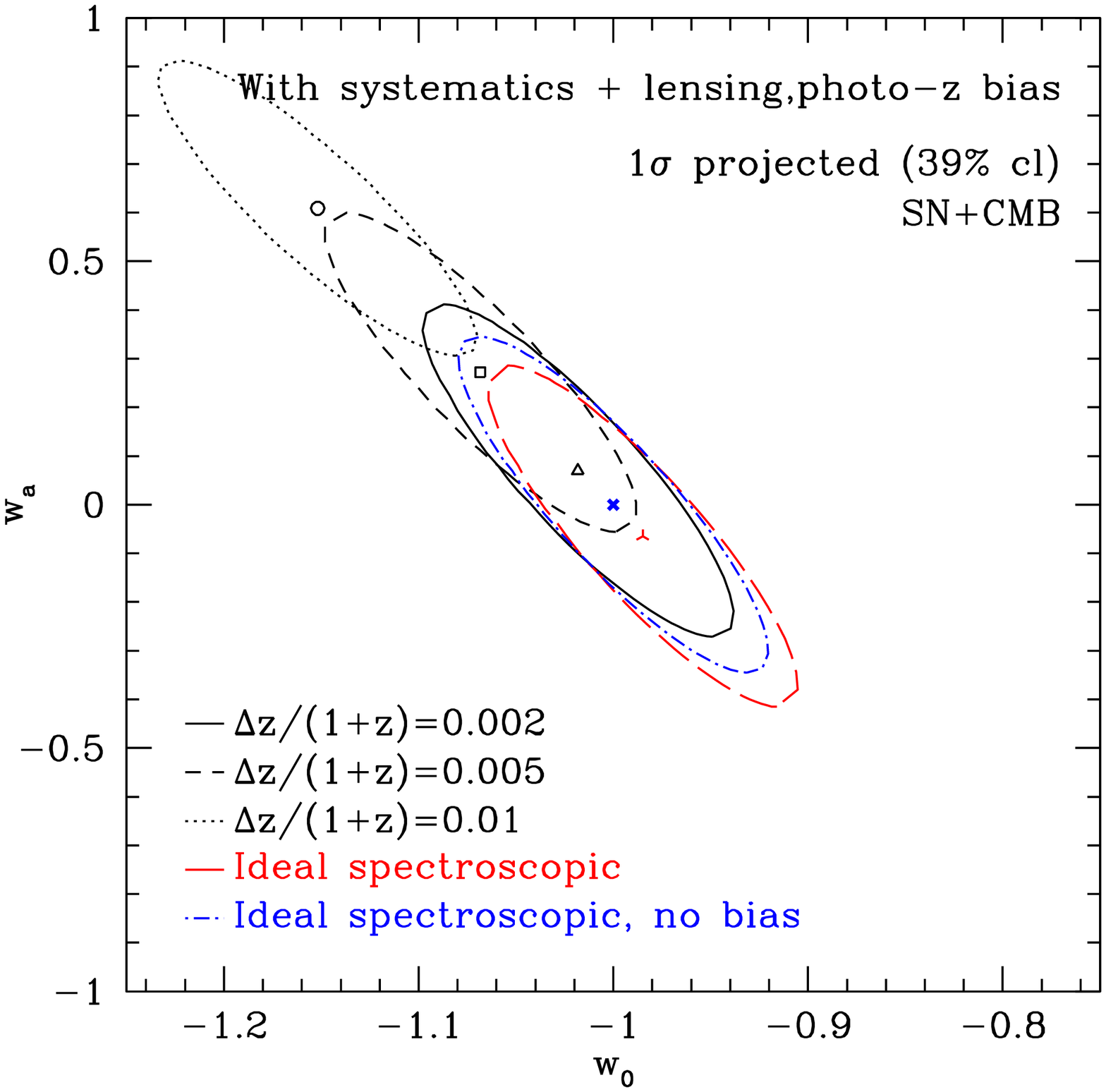, width=3.4in}
\caption{Coherent systematics, such as miscalibrated photometric 
redshifts or incompletely sampled gravitational lensing amplification, 
cause bias in cosmological parameter extraction.  With many sources, 
lensing causes a shift of only a small fraction of the statistical 
uncertainty, from the true cosmology marked with a heavy x to the 
three legged symbol. Photometric redshift errors can give 
biases of a substantial fraction of $1\sigma$ (open symbols). 
}
\label{fig:lenszbias}
\end{center}
\end{figure}

The bias can be quite pernicious as it can mimic a smooth change in 
cosmology (see \citet{bias} for general discussion of biased cosmology).  
Figure~\ref{fig:lenszbias} shows the effect of different 
levels of coherent redshift systematic on cosmological parameter 
misestimation, for a survey extended to $z=3$.  The confidence 
contours are shifted, biasing the 
parameters by amounts visible as the difference between the symbols 
giving the best fit and the heavy x showing the true cosmology. 
We show 39\% confidence level contours so the biases $\Delta w_0$ and 
$\Delta w_a$ can be read directly by projecting to the axes.

The case labeled ``ideal spectroscopic, no bias'' shows the contour 
if one could extend the survey spectroscopically to $z=3$, and 
unrealistically ignoring lensing bias.  The ``ideal spectroscopic'' 
(long dashed, red) contour includes the effect of lensing bias.  We 
see that bias from incompletely sampling the lensing amplification 
distribution is not a major effect, amounting to $<0.2\sigma$ (our 
sample has about 75 sources per 0.1 redshift bin above $z=1.7$; if 
we have fewer sources, the bias goes up but so does the statistical 
uncertainty, so bias is never dominant in the lensing case\footnote{See 
\citet{holzlin} for more discussion of lensing bias, and in particular a 
Monte Carlo treatment that shows the effect on the cosmology contours 
to be asymmetric.  Also see \citet{oguritakahashi} for lensing bias applied 
to gamma ray bursts.}).  

Considering three levels of redshift error, the bias from these 
{\it can\/} be important, shifting the best fit cosmology by 
a significant fraction of the statistical uncertainty, as shown by the 
open symbols.  A reasonable fit to the cosmological bias from the 
propagated redshift error (including stretch) is 
\beq 
\frac{\Delta p}{\sigma(p)}\approx 
0.2\left(\frac{\Delta z}{1+z}\Big/0.001\right), 
\eeq 
for $p=w_0,\,w_a$ (the effect is about 2/3 this on $\Omega_m$).  
Thus the photometric redshifts must be calibrated to 
$\Delta z/(1+z)\lesssim0.002$ if sources at $z>1.7$ are to be used 
for cosmology. 

\begin{figure}[!ht]
\begin{center}
\psfig{file=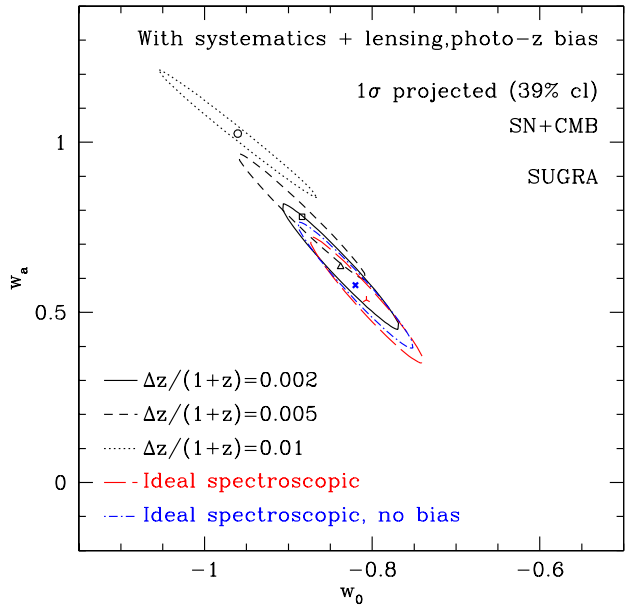, width=3.4in}
\caption{As Fig.~\ref{fig:lenszbias}, for SUGRA dark energy.  While the 
persistence of dark energy density to higher redshifts allows tighter 
constraint of the cosmological parameters, it also makes them more 
susceptible to bias from miscalibrated photometric redshifts. 
}
\label{fig:lenszbiassug}
\end{center}
\end{figure}

While for SUGRA the possible improvement of going from $z=1.7$ to 3 
was $\sim$35\%, we see that the perils of coherent redshift uncertainties  
are also more severe (Fig.~\ref{fig:lenszbiassug}), with parameter 
biases of 0.26, 0.22, and 0.32$\sigma$ respectively on $w_a$, $w_0$, 
and $\Omega_m$ per $[\Delta z/(1+z)]/0.001$.

Overall, the random scatter in photo-z's is not a worry, but possible 
systematic uncertainties require calibration at the challenging level 
of 0.002 if the high redshift objects are to give accurate cosmological 
constraints.  Spectroscopy of a thousand $z>1.7$ SN would be quite 
expensive in time required.  In any case, we emphasize that, apart from 
the numerous pitfalls, standardized candle surveys at $z>1.7$ offer 
little cosmological leverage on dark energy.

\section{Measuring supernovae beyond \lowercase{$z=1.7$}} \label{sec:measure}

Very high redshift supernovae can be quite interesting for astrophysical 
issues (which in turn might impact cosmology), to be discussed in 
\S\ref{sec:sys}.  Here we consider the 
measurement aspects of such a sample, and the implementation of 
observations characterizing the sources, as a practical counterpoint 
to the theoretical considerations of the previous section. 

\subsection{Rates} \label{sec:rates} 

\begin{figure}[!ht]
\begin{center}
\psfig{file=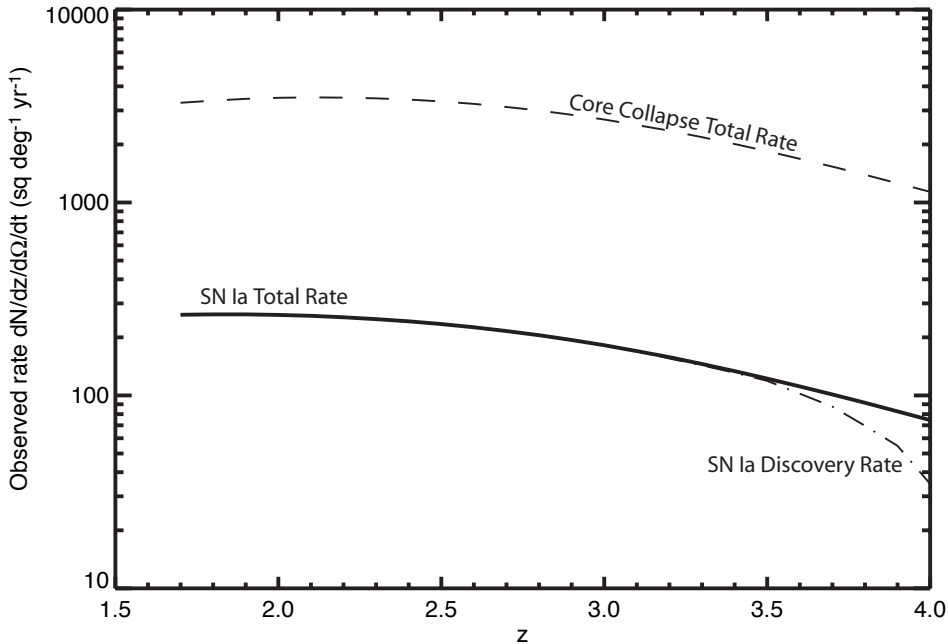, width=5.in}
\caption{The solid and dashed curves show the expected intrinsic
observer-frame supernova rates $dN/(dt d\Omega dz)$ of Type~Ia and
core-collapse supernovae respectively, based on the models discussed
in the text.  The dot-dashed line is the expected discovery rate of SN~Ia
with SNAP, adopting an AB=28 limiting magnitude. While uncertainty
in the luminosity function precludes a straight-forward prediction of
the detectability of the highest redshift core collapse SNe, the SNAP
detection threshold should cause a significant roll-over by $z\sim2-2.5$. 
}
\label{fig:ratesn}
\end{center}
\end{figure}

Thousands of supernovae should exist, and be detected by SNAP, at $z>1.7$.
See Figure~\ref{fig:ratesn} for estimates of both rates; the intrinsic
SNe~Ia rates are from a fit of observed Supernova Legacy Survey (SNLS:
\citet{astier06}) rates at $0.3 < z <0.8$ \citep{sullivanrates} to the
model of \citet{scannapieco05b}.  In this model, each of two SN~Ia
populations has rates proportional either to the star formation rate
(SFR) or the total stellar mass.  Core-collapse supernova rates are based
on HST GOODS rates at $0.3 < z < 0.7$ \citep{dahlen04} and assume that
supernova rates are proportional to the SFR.

We calculate the efficiency of SNAP discovering SNe~Ia by assuming a
limiting AB magnitude of 28 in each of its nine passbands \citep{aldering04}, 
and a 0.35 magnitude dispersion before correction for light curve shape.
We find that SNAP will be almost complete in its discoveries out to $z=3.5$.

For core-collapse supernovae, their heterogeneity and the unknown relative 
rates of subtypes prevent a meaningful calculation of expected discoveries 
relative to the total rate shown in Fig.~\ref{fig:ratesn}.  
Instead, SNAP itself will furnish unique information on the core-collapse 
population at high redshift.  SNAP's depth will provide high-redshift 
core-collapse rates and probe deep into the faint end of their 
luminosity functions.  

The measured rate of core collapse supernovae as a function of redshift
will have important implications outside of stellar astrophysics.
For one thing, SN feedback is an important ingredient in galaxy
formation and properties (e.g.\ \citet{dekel86,scannapieco05a}).
For another, it determines the spectrum of the SN background in neutrino
flux calculations.  The Super-Kamiokande detector, once gadolinium
trichloride is added to its water, will have excellent sensitivity to the
SN background neutrino flux \citep{beacom04}.  Since the threshold will
be $\sim$ 10 MeV (comparable to the average neutrino energy expected from
core collapse SNe), the measurable neutrino flux will have contributions
from SNe up to $z\ls2$ \citep{ando04}.  A well measured core collapse
supernova rate up to high redshifts would thus allow unfolding the
average SN energy spectrum \citep{yueksel05}.

\subsection{Light curves} \label{sec:lightcurves} 

Extracting supernova science from the detected sources requires good 
characterization of their light curves, or flux vs.\ time behavior. 
We calculate the expected signal-to-noise ($S/N$) of the multi-band light
curves generated by the ``deep'' SNAP survey, which repeatedly scans the 
same sky area, using the SNe~Ia spectral template library of \citet{nugentweb}. 

Even for a $z=3$ SN~Ia that is 0.75 mag fainter than normal, we determine
that in the reddest SNAP filter there will be 4 epochs with $S/N > 5$
(and many more at a frequency of one point per rest-frame day with
slightly lower $S/N$).  SNAP will be complete for at least 98\% of the
unextincted, unlensed supernova luminosity function and essentially
complete for $s=1$ SNe~Ia.  Malmquist bias will be negligible.
Section~\ref{sec:lenstype} discusses the effect of lensing amplification
on the SN discovery rate.

From the projected light curve S/N we can estimate how well
the SNe~Ia distances can be determined.  The SNAP filter set consists of nine
filters evenly spaced in $\log{\lambda}$ where the effective wavelength
of filter $n$ is $\lambda_n=(1.16)^n\times4400$\AA\ with $n\in\{0,\dots 8\}$.
This spacing, somewhat finer than that of the Johnson-Cousins set,
bounds $B$ to $V$-band K-correction uncertainties to less than 0.02~mag
\citep{davis06}.  The supernova-frame $B$-band shifts out of the
penultimate $n=7$ filter at $z \gtrsim 1.8$. The low UV flux gives
a weak supernova signal in the $n=7$ filter at $z \gtrsim 3$.  Thus,
color-corrected supernovae at $1.8<z<3$ can be used to measure distances
if SNe~Ia are standardizable over rest-frame $2500-4000$\AA\ (i.e.\ around 
$U$ band).

Whether the $U$-band is useful for determining supernova distances is an
open question.  Theory suggests that the UV emission is sensitive to
metallicity \citep{dominguez01,lentz00}.
\citet{jha06} find that $U$ peak magnitudes are not tightly
standardizable, but show a correlation between optical and $U$ light-curve
decline rates.  \citet{guy2005} have used published photometry of low-$z$
SNe~Ia to establish that $U$-band is stable enough to use for
color/extinction corrections.  Using the same SN light-curve modeling
technique, \citet{astier06} show that the $U$ magnitude can be predicted
using $B$ and $V$ --- to within $\pm0.03$~magnitude for well-measured
SNLS supernovae. In addition high-quality restframe UV spectroscopy of SNLS
SNe~Ia shows good agreement with that of low-redshift SNe~Ia
\citep{sullivantbd}.

SNAP itself provides UV peak-brightness spectroscopy and light curves
of thousands of $z<1.7$ supernovae.  If SNe~Ia are indeed
standardizable in the UV, these data will be used to derive
distance-determination algorithms for even higher redshift supernovae.
The UV spectral template can be constructed from the subset of objects
with measured spectral time-evolution and the supplementary coarse
multi-epoch spectroscopy derived from redshift-dependent
light-curve shapes.  K-corrections and uncertainties for the off-peak
photometry necessary for analysis of photometry lacking a
corresponding spectrum can be determined as in \citet{nugent02,davis06}.

Assuming UV standardizability, the SNAP wavelength window provides
sufficient coverage to allow extinction corrections of $z>1.7$
supernovae.  Based on the expected $S/N$ of $z=2$, 2.5, and 3
supernovae, we calculate the uncertainty in distance modulus $\mu$ 
obtained from a simultaneous fit of the \citet{ccm} dust-model
parameters $A_V$ and $R_V$ as well as the supernovae parameters of
stretch, peak magnitude, and time of explosion, using the Fisher
matrix formalism.  For a $z=2$ supernova we find $\sigma(\mu)=0.15$
and 0.05 while fitting for the two- and one-parameter ($R_V$ fixed) 
dust models respectively, compared to the intrinsic dispersion 
$\sigma_{\rm int}\approx0.15$.  However, for $z=2.5$ these numbers increase 
to 0.52 and 0.09, 
while for $z=3$ the two-parameter model is unconstrained, and when 
$R_V$ is fixed to 3.1 then $\sigma(\mu)=0.16$.  Uncertainty in the 
wavelength dependent dust absorption is a major systematic for supernova 
cosmology.  SNAP is specifically
designed to control possible dust evolution for SNe at $z<1.7$ by 
allowing fits of multiparameter dust models.
However, SNAP can furnish standardized distances for  SNe out to $z=3$ if
the dust properties can be adequately constrained. 

The expected achievable precision per supernova and the large 
sample (see \S\ref{sec:rates}) with high S/N measurements 
allow determination of the average 
magnitude in a 0.1 redshift bin to about 0.005 mag.  
Since the impact on cosmology of an individual bin is small and 
strongly correlated, one can use bin-to-bin variations to constrain 
or identify potential systematic effects from K-corrections and photo-z's.

\subsection{Redshift determination} \label{sec:photoz}

The calculations of \S\ref{sec:zerr} showed the importance of accurate
redshifts for the high-redshift Hubble diagram.  In addition,
knowledge of redshifts aids in selecting targets for follow-up
complementary to that which SNAP will obtain on its own.  Here we
discuss implementation of such measurements.  Obtaining redshifts for
such high-redshift SNe will pose a challenge, due to both their
faintness and transient nature.  Measurement of photometric redshifts
of the SN host galaxies can provide an efficient, and possibly
necessary, alternative, in combination with calibrating spectroscopic
redshifts from the James Webb Space Telescope (JWST) or a thirty meter
telescope (generically TMT).  We discuss some pros and cons of this
and some other possibilities in the remainder of this section.

Supernova host galaxies are expected to have bright luminosities.
Even more so than for local SNe~Ia the production rate at high redshift 
should follow the rest-frame $B$ 
luminosity, and thus the distribution of host galaxy luminosities
tracks the luminosity-weighted galaxy luminosity function. The
incidence of core-collapse SNe tracks star formation, and star-forming
regions have high surface brightness and emit across the full range of
SNAP filters.  In the SNAP deep field it should be possible to obtain
$S/N=10$ photometry for (compact) galaxies as faint as $m_{AB}\sim29$
at optical/NIR wavelengths \citep{aldering04}, corresponding to
$M_B\sim-18$ at $z=3$.  SNAP is thus able to provide precision
multi-band photometry covering the spectral features necessary for
photometric redshift determination of a large fraction of supernova
hosts out to $z=3$.

Spectroscopic redshifts are required to accurately calibrate SNAP
photometric redshifts.  The accuracy of SNAP photometric redshifts of
galaxies during the mission should be sufficient to allow reliable
selection of SNAP transients for complementary follow-up studies
\citep{aldering04,dahlen06}. The final set of precise photometric
redshift calibration necessary for cosmological analysis would not be
needed until SNAP observations are well underway.  Even for
photometric redshifts with scatter at very high redshift of
$\sigma_{\ln(1+z)}\sim0.05$, one thousand calibration spectra could
achieve the level $\Delta z/(1+z)\sim0.002$ needed to use SNe~Ia with
$z>1.7$ for unbiased cosmology (see \S\ref{sec:zerr}).  (Recall that
for $z<1.7$, $\partial m/\partial z$ is larger and significantly more
stringent redshift determination is required.)

The necessary calibration reaching to the photometric redshift limit
of SNAP is probably tractable with multi-slit spectroscopy of galaxies
in the SNAP deep field using JWST or TMT.  The surface density of
galaxies in the SNAP deep field will be of order 250~arcmin$^{-2}$
\citep{aldering04}, allowing for efficient multi-object spectroscopy.
At optical wavelengths TMT could easily obtain redshifts to
$m_{AB}=26$ for a few thousand galaxies per night; this would be
sufficient to calibrate the brighter star-forming galaxies. For the
oldest --- UV-faint --- galaxies, near infrared (NIR) spectroscopy
with NIRspec on JWST might be necessary.  Self-contained spectroscopy
of galaxies which land ``for free'' in the SNAP IFU spectrograph ---
obtained in parallel with SNAP imaging observations --- also provides
some calibration of photometric redshifts.  NIRspec on JWST would
require 30~hours to reach $S/N\sim10$ per resolution element for
m$_{AB}\sim28$ galaxies with $R\sim100$, and would be able to observe
$\sim100$ objects simultaneously at NIR wavelengths
\citep{jakobsen06}. TMT would require of order 80~hrs to reach
comparable $S/N$ to m$_{AB}\sim28$ at optical wavelengths, with
$\sim500$ galaxies observed at once.  Multi-week campaigns on these
facilities would be necessary to achieve sample sizes of order 1000
galaxies with m$_{AB}\sim28$. Note that follow-up of the SNAP deep
fields is likely to be of major interest for, e.g., JWST and TMT,
independent of a need for photometric redshift calibration.

The efficacy of measuring photometric redshifts from the supernovae 
themselves depends heavily on SN physical properties.  Type Ia 
supernova light curves, due to
their phase-dependent color evolution and sensitivity to broad
spectral features, enable photometric determination of redshifts.
Current algorithms and templates applied to the SNLS dataset predict
photometric redshifts for which 90\% of SNe~Ia deviate by $\ls 0.08$
and the median absolute error is 0.03 when compared to
spectroscopically measured redshifts \citep{sullivan06}.  Next
generation telescope surveys with more filters and higher $S/N$ should
yield further improvement.

The precision and accuracy that supernova photometric redshifts can
achieve are limited, however.  Complications could arise if the SN
photospheric velocity, which decreases as the light curve brightens
and fades, were to change systematically with redshift. As shown in
\S\ref{sec:zerr}, a photometric redshift bias no larger than 0.002 can
be tolerated for $z>1.7$ cosmological measurements.  This translates
into a maximum allowable shift in the ensemble photospheric velocity
--- potentially introduced by sample selection biases or intrinsic
changes in SNe~Ia --- of 600 km/s. The scatter amongst SNe~Ia at a
given phase is $\sim1300$ km/s
\citep{benetti05}. Hence the ensemble velocity need only shift by less
than half the observed scatter in order to greatly impact cosmological
measurements at such high redshift. Thus external constraints on
expansion velocity evolution, e.g.\ up to $z=1.7$ from the SNAP
spectrograph, will be needed to assess the efficacy of using
photometric redshifts from the SNe~Ia themselves for cosmology.

As described above, calibrated photometric redshifts from the host
galaxies are sufficient to satisfy the bias limit for the redshift
range $z>1.7$. Independently determined photometric redshifts from the
SN lightcurves could then play a role in resolving remaining
ambiguities. The accuracy of the photometric redshifts is certainly
sufficient for triggering targeted follow-up programs, e.g.\ using JWST.

\subsection{Supernova typing} \label{sec:typing}

Categorizing the supernovae into types is an essential element of 
supernova, astrophysical, and cosmological studies. 
Photometric segregation of supernovae can be performed with
some success based on lightcurve and color evolution. Early
variations of this approach have been presented in
\citet{poznanski02,galyam04,riess04,barris04}. In general, when
restframe UV data are available Type~II SNe are seen to be UV-bright
whereas opacity from iron-group elements in SNe~Ia suppresses the UV
brightness. In addition, the lightcurves of SNe~IIP are quite distinct
from those of other SN types.  
Using a Monte Carlo lightcurve simulation
appropriate to the color coverage, cadence, and depth of SNAP we confirm
that for the $S/N$ achievable for SNe~Ia at $z=3$ the shape of the
B-band lightcurve distinguishes between Type~Ia and Type~IIP SNe.  

Distinguishing Types Ib and Ic from Type Ia is more difficult,
especially in the face of an uncertain redshift and dust
extinction. SNe~Ib/c are generally redder than SNe~Ia (with restframe
B-V color about 0.5 magnitudes redder \citep{poznanski02}). The
color evolution is different as well: Type Ib/c have similar pre-
and post-maximum colors while Type Ia become redder after their
maximum brightness is reached. The colors and magnitude can be used 
to largely break the
degeneracy between dust reddening and SN type once a full lightcurve
is obtained. The precise multi-band photometry afforded by SNAP can
greatly improve the power of such techniques.

The largest and most complete sample on which this technique has been
tested to date is the Supernova Legacy Survey. \citet{sullivan06} use 4-color
lightcurve photometry to reject all ten SNe~II while rejecting only one
SN~Ia in a sample of 85 spectroscopically-classified high-redshift SNe. 
However, \citet{sullivan06} do not demonstrate their ability to reject 
SNe~Ib/c and they do not comment on whether or not all the SNe~II were IIP,
or might have included SNe~IIL.

At the redshifts $z>1.7$ considered here, all stellar populations will be 
young, and therefore
galaxy morphology based shortcuts applicable at lower redshift, e.g.\ that
ellipticals never host core-collapse SNe, cannot be used. With the spatial
resolution (e.g.\ 1~kpc at $z=3$) and multicolor information from SNAP it
may be possible to sufficiently constrain the stellar population age at
the (projected) location of a SN to partly discriminate some core-collapse
SNe from SNe~Ia. The critical assumption here is that most core-collapse
SNe will be produced in regions dominated by a single starburst, and that
the light of such a starburst will have high surface brightness and thus
dominate over any underlying older population. In this case multicolor
photometry covering the restframe UV can constrain the starburst age and
therefore the lowest mass of the stars completing their evolution. Taking
$8~M_\odot$ as the dividing line between core-collapse and thermonuclear
supernovae \citep{iben83}, evidence of a starburst younger than 50~Myr
\citep{schaller92,schaerer93} at the location of a SN would strongly
suggest that the SN is a core-collapse SN. Protracted starbursts,
low-level star formation, and projection effects will complicate this
technique. Absence of detectable star formation anywhere near a SN would
suggest a SN~Ia, but it will be difficult to rule-out low-level star
formation and hence a core-collapse origin.  Again, SNAP spectroscopic
follow-up of SNe hosts at lower redshift will help calibrate this technique.

Detection of the shock breakout from core-collapse SNe would provide
a completely new and independent means of discriminating SN types. 
As discussed in \S\ref{sec:corecoll}, predictions of 
luminosity and timescale, and hence
detectability, are quite uncertain. It is encouraging in this respect that
shock breakout has been detected at restframe optical wavelengths in low
redshift Type II (SN~1987A, \citet{menzies87,hamuy88}), IIb (SN~1993J,
\citet{lewis94,richmond94}), and Ib/c (SN~1999ex, \citet{stritzinger}) SNe
with timescales and luminosities that SNAP could detect out to high
redshift. SNAP will sample the restframe UV of high-redshift SNe,
where the hot shock breakout will be brighter, however, the fastest
events may be missed due to SNAP's 1-2 day restframe cadence. 

It is tempting to exploit the fact that core-collapse SNe 
are generally fainter at peak than SNe~Ia to obtain the type. 
However, the luminosity functions do overlap 
\citep{richardson02,richardson06}, and gravitational 
lensing (see the next section), extinction, and redshift errors can further 
blur this distinction.  Using luminosity as the sole type 
discriminant would surely bias any cosmological measurement 
by modifying the statistical distribution of true SNe~Ia and 
inviting interlopers \citep{homeier05}. Luminosity still may 
prove useful as a weak prior in concert with other type 
discriminants. 
Without the need to 
trigger spectroscopic followup, one can wait for the full light curve 
to type the supernova.  The magnitude might, however, be used to feed 
the supernova to JWST or TMT; this poses no danger as that 
spectroscopic information will confirm the type.

\subsection{Typing and lensing amplification} \label{sec:lenstype} 

Gravitational lensing can confuse typing through both amplification 
and deamplification of the intrinsic luminosity.  
While SN~Ib/c have a wider intrinsic dispersion
than Ia, typically they are 1-2.5 magnitudes fainter. 
The probability of SN~Ib/c being sufficiently strongly amplified by
lensing to reach Ia luminosity is small, and likely would be accompanied
by multiple images or other signs that lensing is present.

We can assess the converse process -- Ia being demagnified by 1 mag -- 
through looking at the maximum demagnification possible as a function 
of redshift.  Since magnification is dominated by the Ricci focusing 
of light by density inhomogeneities, and one cannot have an underdensity 
with density less than zero, then the maximum demagnification is determined 
by the empty beam distance $d_{\rm empty}$ where all matter is removed 
from the line of 
sight.  The distance-redshift relation for arbitrary clumpiness 
(over- or underdensity) and arbitrary dark energy equation of state 
was derived by \citet{lin88}.  The demagnification is then bounded 
by $\Delta m<5\log(d_{\rm empty}/d)$, where $d$ is usual 
Friedmann-Robertson-Walker luminosity distance. 

\begin{figure}[!ht]
\begin{center}
\psfig{file=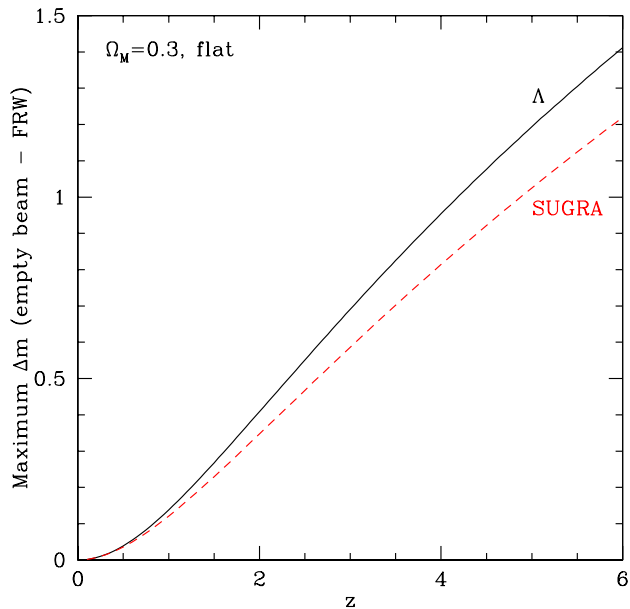, width=3.4in}
\caption{The maximum demagnification is given by difference between 
apparent magnitude in an empty beam model (zero matter along the line 
of sight) and a smooth, FRW model.  The curves show the results for 
two different dark energy models. 
}
\label{fig:demag}
\end{center}
\end{figure}

Figure~\ref{fig:demag} shows that for $z<3$ the demagnification is 
less than 0.7 mag and hence SNe~Ia are only 
likely to overlap with the bright tail 
of the SNe~Ib/c luminosity function in the most extreme empty 
beam case, and only at the highest redshifts. 
A fitting formula valid to better than 0.01 mag 
out to $z=4$ for the maximum demagnification in a dark energy cosmology 
is given by 
\beqa 
\Delta m&=&Az^2,\quad z<1 \\ 
&=&A+1.93A(z-1),\quad z\ge1 \label{eq:hizminlens} 
\eeqa 
where $A=0.14-0.04[1+w(z=1)]$.

\subsection{Using lensed high redshift supernovae} \label{sec:slens} 

In addition to acting as noise, causing dispersion and bias, gravitational 
lensing also provides an astrophysically useful signal.  The extension 
of Eq.~(\ref{eq:siglens}) to high redshift contains important 
information on structure formation.  The dispersion as a function of 
depth can provide constraints on the mass amplitude $\sigma_8$ and the 
evolution of the matter power spectrum \citep{frieman97,dodelson}, 
while the cosmic variance along different lines of sight also carries 
valuable knowledge of the matter distribution \citep{pen04,cooray06}. 

Individual, strongly lensed standardized candles, as well as their 
statistics, provide important probes of cosmology and galaxy cluster 
mass profiles and velocity dispersions through the amplification time 
variation, multiple image separations, image flux ratios, and time 
delays \citep{holz01,goobar02,lewisibata,linsl}.  The high quality, 
multiband, cadenced sample of tens of strongly lensed, very high 
redshift supernovae from SNAP will contribute a rich science resource.

\section{Astrophysics and systematics} \label{sec:sys} 

Very high redshift supernovae open windows on several areas of 
astrophysics.  We examine here the use of such observations to 
learn about supernova physics and dust properties.

\subsection{Progenitor and environmental systematics} \label{sec:sys17} 


The study of supernovae at high redshifts offers the possibility of
seeing evolutionary effects in action. SNAP will exploit this feature
extensively to decouple systematics from the true cosmological signal
using its $z<1.7$ sample.  Given the relative insensitivity of luminosity
distance to dark energy for $z>1.7$ (see \S\ref{sec:dark}), deviations in
the Hubble diagram could potentially be useful as a means to reveal 
astrophysical systematics \citep{riess06}.  The value of such higher redshift 
SNe~Ia for probing systematics that could affect cosmological measurements
thus hinges on whether or not some property of SNe~Ia is especially-well
probed or constrained in this redshift interval. Such circumstances might
include extraordinarily low metallicity or extension of the lookback time
early enough such that the upper limit set by the age of the universe
meaningfully constrains the timescales for SN~Ia progenitor models.

In reality, extending the study of SNe~Ia from $z=1.7$ to $z=3$ increases
the lookback time by a mere 1.7~Gyr.  Since the age for a $\om=0.3$, flat
$\Lambda$CDM universe at $z=3$ is 2.2~Gyr, the first generations of stars
in the canonical 3--8~$M_\odot$ mass range expected for C/O white dwarf 
progenitors
\citep{iben83} will have evolved and begun producing SNe~Ia via the
various channels currently in contention \citep{belczynski05}. Therefore,
it will remain difficult to directly probe systematic effects due to
the progenitor mass. Even proposed channels with long delay times of
2--4~Gyr \citep{strolger04,scannapieco05b} would be operative over the
$1.7<z<3$ redshift range.  Therefore, in order to distinguish between
progenitor scenarios, detailed modeling of star formation histories,
binary parameters, etc., for various progenitor scenarios will be required
just as it is for the $z<1.7$ SN~Ia population \citep[e.g.][]{forster06}.

Similarly, one cannot 
expect SNe~Ia progenitors to be dominated by low-metallicity systems.
Most of the gas that assembles into stars that will
explode as SNe~Ia at $1.7<z<3$ will be enriched with metals,
as the most massive galaxies are already in place by $z\sim3$
\citep{steidel96,heavens04,vandokkum06}. Even the lower-density, lower
SN-yield, environments probed by damped Lyman-$\alpha$ absorbers are
enriched. This can be seen directly from [Zn/H] measurements of 
damped Lyman alpha systems (DLAs) 
\citep{prochaska03,kulkarni05}, which show $-0.4<{\rm [Fe/H]}<-2.5$
over the range $1.7<z<4$. This is despite the fact that DLAs will
exhibit lower metallicity than the average due to several selection
effects, including the preferential selection of outer regions of the
galaxies associated with the DLAs, metallicity gradients in galaxies,
the upper limit on $N_{HI}$ included in the definition of DLAs, etc.\ 
\citep{zwaan05,johansson06}. This range in DLA metallicity is not that
much larger than that found amongst nearby galaxies. 

Determining the evolution timescale(s) is important for understanding the
allowed combinations of binary systems leading to the creation of SNe~Ia.
The ideal laboratory for such studies is a delta-function episode of
star-formation, since then the starting time is defined. In reality one
must deal with finite star-formation timescales, either within specific
galaxies or across all galaxies.  The rising global SFR at $z\sim3$
\citep{hopkins06} means that those progenitor systems formed at much
higher redshift with potentially much lower metallicity and having a
long delay time will be overwhelmed in numbers by younger progenitors
having lower delay times. Conversely, the dramatic roll-over of the
global SFR for galaxies at $z<1$ \citep{heavens04,juneau05,hopkins06}
may make this redshift range the most useful in revealing details of a
progenitor population having a long delay time.

Important effects, especially those related to metallicity, may
still be revealed. For instance modeling by \citet{dominguez01} and
\citet{lentz00} suggest that changes in progenitor metallicity will
affect the metallicity of the ejecta composition, and thus its opacity
and brightness in the UV. On the other hand, \citet{kobayashi98}
have suggested that the production of SNe~Ia will be inhibited as the
metallicity of the gas from the donor star decreases, choking-off around
[Fe/H]~$\sim-1$. The broad ramification of this model for cosmology
is that SNe~Ia from [Fe/H]~$<-1$ progenitors will not exist, and
therefore metallicity-dependent effects will be truncated.  Indeed, as
metallicity correlates much more strongly with galaxy luminosity than
with redshift, metallicity-dependent systematics studies from higher
redshift SNe~Ia may be secondary to what can be learned by the study of
SNe~Ia in local galaxies \citep{hamuy00,gallagher05}.

In the simple galaxy chemical evolution models presented by
\citet{kobayashi98}, the SN~Ia rate would plummet around $z\sim1.2$,
where the galaxy metallicity crosses the [Fe/H]~$\sim-1$ threshold. In
reality the redshift at which a galaxy passes the [Fe/H]~$\sim-1$
threshold will be mass-dependent, and environmental affects on galaxy
star formation histories will further blur the [Fe/H]~$\sim-1$ threshold
in redshift space. Therefore, what may be observed is a gradual drop in
the SNe~Ia rate, superimposed on the modulation already expected due to
the SFR, that could well continue to $z\sim3$. As galaxy metallicities
at $z\sim3$ will be difficult to obtain directly 
(e.g.~\citet{shapley05}), modeling of this effect 
will be strongly coupled to galaxy formation and SF models.

It is also possible that extremely low metallicity environments
\citep{jimenez06}, will lead to so-called ``Type~1.5'' SNe, in which 
the C/O core
of an AGB star reaches the Chandrasekhar limit \citep{iben83}. For
instance, \citet{lowmetalSN} has proposed that ultra-metal-poor
stars, with [Fe/H]~$<-5$ will produce SNe~1.5. It may be possible
that stars up to [Fe/H]~$\sim-1$ can produce SNe~1.5, e.g., if the
mass loss from stellar winds decreases sufficiently with decreasing
metallicity \citep{zijlstra04}.  These events would be hydrogen rich,
possibly similar to SN~2002ic \citep{woo02,hamuy03,woo04} or SN~2005gj
\citep{prieto05,snfactory06}.  The potential of such SNe as standard
candles has yet to be explored.

All things considered, typical $1.7<z<3$ SNe~Ia may not probe a fundamentally
different region of parameter space than those at $z<1.7$. Of course
it is still possible that individual SNe~Ia, e.g.\ in an ultra-low
metallicity or very young environment, will offer important clues to
understanding SNe~Ia.  For these, higher $S/N$ photometry extending to
longer wavelengths, as well as spectroscopy, would be very valuable. In
these cases JWST could be employed if the uniqueness of such SNe can be
realized in time from SNAP observations to trigger dedicated follow-up.
Meanwhile, local SN observations are rapidly expanding efforts
to probe young stellar populations in low metallicity environments
\citep{woo04,snfactory06}, as well as continuing the unique study of SNe~Ia
in old and/or metal-rich stellar populations possible at low redshift.

\subsection{Extinction and dust properties} \label{sec:extinct}

Accurate extinction measurements are necessary for the cosmological
use of supernovae as standard candles.  Alternatively, we can view
supernovae as providing a source with known spectral energy
distribution by which we can constrain the extinction properties of dust.  
This also allows tests of the applicability of the \citet{ccm} extinction 
model at observed SN-frame wavelengths (as discussed in 
\S\ref{sec:lightcurves}, SNe~Ia standardizability in the UV can be tested 
with SNAP data itself).  Beyond this, a foundation for models 
of extinction wavelength-dependence may be laid by study of the subset
of $z<1$ SNe~II caught at shock breakout, through deviations from fixed
blackbody color at the Rayleigh-Jeans tail (see \S\ref{sec:corecoll}).  
Additionally, strongly
lensed quasars in the SNAP fields provide differential extinction on
different lines of sight (\citet{falco99}, but see \citet{mcgough05}).  

SNAP observation of supernovae at $z>1.7$ furnishes information on
dust extinction in the UV.  Together with supplemental JWST followup,
we can have measurements of UV-optical or optical-NIR extinction along
many supernova lines of sight.  This extends such astrophysical dust
studies over the great majority of the age of the universe.

One final challenge will be absorption from the intergalactic medium.
Simulations \citep{madau95,meiksin06} and observations \citep{songaila04}
indicate observer-frame $B$-band absorption of $\sim0.1$~mag by $z=2.5$
and 0.3-0.4~mag by $z=3$, largely from hydrogen. Likewise, observer-frame
$V$-band absorption is expected to be $\sim0.1$~mag by $z=3$. Longer
wavelengths relevant for high-redshift SNe~Ia will not be affected by
hydrogen, but the UV spectra of the brightest SNe~II will be affected.
Metal lines could in principle affect longer wavelengths in certain
filters, depending on the sightline.  But on average they should cause
absorption well below 0.01~mag \citep{songaila05}.  For SNe seen through 
DLA systems broadband extinction may not be negligible 
\citep{ellison05,york06}.

\subsection{Shock breakout in core collapse SNe} \label{sec:corecoll}

Core collapse SNe exhibit the signature phenomenon of shock
breakout \citep{colgate68} when the forward shock reaches the stellar
surface. Observation of this event would help type the SN and carries
potentially valuable information on the physical properties of the
progenitor.  The characteristic timescale of the shock breakout
depends on the structure of the progenitor star, with compact blue
supergiants (BSG) expected to produce shocks with cosmological
restframe timescales of 0.03~hrs, compared to 0.3~hrs for the much
more extended red supergiants (RSG) \citep{matznermckee}. These values
for all but the largest stars will be affected 
by the light travel time across the star 
\citep{ensman92}.  Both progenitor types are expected to produce peak
shock temperatures in the range $T\sim0.5-1\times10^6$~K and peak
luminosities of $L\sim10^{45}$~erg~s$^{-1}$. While this emission peaks
at soft X-ray wavelengths, there can be significant emission in the
restframe UV wavelengths that SNAP will sample. Moreover, the timescale
for strong emission in the UV can range 
from a significant fraction of a day to several days
\citep{ensman92,young95,blinnikov98,blinnikov00}. 

There exist four examples of low-redshift core-collapse SNe to date
where detection of shock breakout is certain (SN~1987A, SN~1993J, and
SN~1999ex) or probable (GRB~060218/ SN~2006aj; for this case there is
concern that the emission is not necessarily all from classical shock
breakout). To our knowledge, these also constitute the full sample
of core-collapse SNe discovered within a day or two of explosion,
but we caution that still it is possible that they are not representative.  
Restframe U-band observations of these events indicate approximate 
un-dereddened 
absolute AB magnitudes greater than $-15.3$, $-19.5$, $-14.7$
for SN~1987A, SN~1993J, and SN~1999ex, respectively, and equal
to $-18.2$ for GRB~060218/SN~2006aj during the shock breakout phase
\citep{moreno87,menzies87,hamuy88,lewis94,richmond94,stritzinger,campana06}.
(The lower limits reflect the fact that the peak of shock breakout 
emission in $U$-band was missed in most cases.)  Perhaps more relevant
for typing purposes is that for all cases except SN~1999ex, the shock
breakout phase was the brightest epoch of the $U$-band lightcurve.  Thus,
if SNAP detects a core-collapse SN, it appears likely that it will also
be sensitive to the shock breakout.  

As SNAP will sample further into the restframe UV --- down to 
restframe Ly-$\alpha$ in its bluest bands, detections
should be somewhat stronger, depending on the amount of dust extinction.
For example at restframe 1800~\AA, GRB~060218/SN~2006aj exhibited an
absolute AB magnitude of $-18.4$ \citep{campana06}.  The shock 
breakout for SN~1993J, having a RSG progenitor
\citep{aldering94,vandyke02}, is one example that SNAP should be
capable of detecting out to $z\sim3$. 

SNAP samples any given patch of sky with 4 exposures of 300~sec in
each of the bluest (optical) filters and with 8 exposures of 300~sec
in each of the reddest (NIR) filters. Consecutive observations in
each of the 9 SNAP filters means that a given patch of sky is
monitored continuously for 1.1~hrs every 4~days in the observer
frame.  Thus, with fortuitous timing --- likely for some fraction
of the thousands of core-collapse SNe SNAP will detect --- SNAP may
obtain multi-epoch coverage in up to 9 filters, possibly covering
the most luminous period of an event even for fast-declining BSG
progenitors. This will allow a determination of the luminosity and
temperature (spectral) evolution of the shock breakout. In simple
shock breakout models \citep{matznermckee} this evolution is set
primarily by the stellar radius, followed in importance by the
explosion energy, and finally the total ejecta mass. Determinations
of the redshift and dust extinction will be needed to set the correct
scale for these quantities for each SN.  Some common physical
parameters (such as opacities --- expected to be dominated by
electron scattering --- and characteristic density profiles) will
need to be determined for the population as a whole in order obtain
the best constraints \citep{matznermckee,calzavara04}.  However,
distinguishing between BSG and RSG progenitors for shock breakout
events with good temporal coverage should be straightforward, and
this discrimination is already sufficient to provide more detailed
constraints on the star formation history of the universe than can
be obtained from the integrated light of galaxies or from SN rates
alone.

\section{Conclusions} \label{sec:concl}

As part of its normal operation SNAP will discover and obtain lightcurves
for thousands of SNe with $z>1.7$. For SNe~Ia, we find that SNAP will have
good sensitivity out to $z\sim3$, where the nominal 4-day observer-frame
cadence will equal an unprecedented daily cadence in the rest-frame.

However, we find that attempting to use $z>1.7$ standardized candles
for cosmology leads to weak statistical gains and is prone to new
systematic errors  -- a high risk, low yield strategy that undoes
the careful systematics control SNAP provides at $z<1.7$. We conclude
that in the case of SNe~Ia, limited wavelength coverage, and lack of SN 
and host spectroscopy, would additionally open the door to biases in redshift,
contamination from SNe~IIL and Ib/c, and poorer standardizability due to
dust and metallicity in the rest-frame UV. These observational issues could
be remedied in principle with supporting JWST spectroscopy and rest-frame
optical lightcurves. Largely due to overhead, SNAP-quality follow-up of
even 100 SNe~Ia in the range $1.7<z<3$ with JWST would constitute a very
large program, and attaining SNAP-level calibration might also be an issue.

While the utility of such SNe~Ia appears marginal for precision measurements
of the cosmological parameters, such SNe --- both thermonuclear and
core collapse --- will be valuable for understanding many aspects of
stellar and galaxy evolution. The core collapse SNe will further help
in understanding the neutrino background. JWST follow-up of the most
unusual SNe, including lensed SNe, could also prove useful --- for general astrophysical 
problems, and perhaps for dark matter and cosmological measurements.

\acknowledgments

We thank Peter Nugent for helpful discussions.  
This work has been supported in part by the Director, Office of Science,
Department of Energy under grant DE-AC02-05CH11231.

\end{document}